\documentclass[aps,prd,superscriptaddress]{revtex4}

\usepackage{amsfonts}
\usepackage{amsmath}
\usepackage{array}
\usepackage{xcolor}
\usepackage{graphicx}
\usepackage{color}
\usepackage{amssymb}
\usepackage{esint}
\usepackage{empheq}
\usepackage{mathtools,leftidx}
\usepackage{float}
\usepackage[T1]{fontenc}
\usepackage{bbold}

\usepackage{soul}

\DeclareMathAlphabet\mathbfcal{OMS}{cmsy}{b}{n}

\begin{document}

\title{Stable Magnetic Lorentz-Violating Vacua in Gauge-Invariant Nonlinear Electrodynamics}

\author{E. Pl\'acido-Flores}
\email{lalo@xanum.uam.mx}
\affiliation{Departamento  de  F\'isica,  Universidad  Aut\'onoma  Metropolitana-Iztapalapa, San Rafael Atlixco 186, 09340 Ciudad de M\'{e}xico, M\'{e}xico}

\author{Rom\'an Linares}
\email{lirr@xanum.uam.mx}
\affiliation{Departamento  de  F\'isica,  Universidad  Aut\'onoma  Metropolitana-Iztapalapa, San Rafael Atlixco 186, 09340 Ciudad de M\'{e}xico, M\'{e}xico}

\author{V.  L\'opez}
\email{cbi2193054261@izt.uam.mx}
\affiliation{Departamento  de  F\'isica,  Universidad  Aut\'onoma  Metropolitana-Iztapalapa, San Rafael Atlixco 186, 09340 Ciudad de M\'{e}xico, M\'{e}xico}

\author{C. A. Escobar}
\email{carlos.escobar@xanum.uam.mx}
\affiliation{Departamento  de  F\'isica,  Universidad  Aut\'onoma  Metropolitana-Iztapalapa, San Rafael Atlixco 186, 09340 Ciudad de M\'{e}xico, M\'{e}xico}

\begin{abstract}
We investigate gauge-invariant nonlinear electrodynamics in the Pleba\'nski first-order Hamiltonian formulation, taking the single-invariant potential $\hat V(P)$ as the primary object. Our focus is on the existence of stable Lorentz-violating magnetic vacua. For three explicit two-parameter models---rational asymmetric, logarithmic, and exponential---we determine the regions of parameter space in which nontrivial constant electromagnetic vacua are compatible with an effective Hamiltonian bounded from below and a positive-semidefinite Hessian. In all three cases, physically admissible Lorentz-violating vacua are realized in the magnetic branch. We further discuss the electric branch and several additional one-parameter models, illustrating that Hamiltonian boundedness by itself does not ensure spontaneous Lorentz symmetry breaking. We also comment on how the symmetry-breaking conditions are related to known strong-field causality criteria.
\end{abstract}

\maketitle

\section{Introduction}

Lorentz invariance is one of the fundamental symmetries underlying both the Standard Model of particle physics and general relativity. Yet a number of approaches to quantum gravity suggest that it may not be exact, and that departures from exact Lorentz symmetry could arise 
in an underlying fundamental theory, either explicitly or through spontaneous symmetry breaking \cite{KosteleckySamuel1989,Amelino}. The Standard-Model Extension (SME) provides a general effective-field-theory framework in which such effects can be described in a model-independent manner \cite{Kostelecky,Kostelecky2}. In many SME scenarios, the Lorentz-violating coefficients are understood as vacuum expectation values of underlying dynamical fields.

Spontaneous Lorentz symmetry breaking has been extensively studied in models in which vector or tensor fields acquire nonzero vacuum expectation values. In electrodynamics, a well-known example is provided by bumblebee models, where a vector field develops a nontrivial vacuum value through a symmetry-breaking potential, thereby inducing Lorentz violation, typically at the cost of gauge invariance \cite{Hernaski1,Escobar1}. Although such constructions are conceptually appealing, they often require a careful restriction of phase space in order to preserve stability and avoid negative-energy excitations \cite{Potting1}. More generally, they illustrate that the realization of Lorentz-violating vacua is closely tied to the dynamical consistency of the theory.

The role of spontaneous Lorentz breaking in gravitational and SME contexts has also been emphasized in Ref.~\cite{Kostelecky2004}, where explicit Lorentz breaking was shown to be incompatible with generic Riemann--Cartan geometries, while spontaneous breaking avoids this obstruction. 

Related realizations include antisymmetric-tensor models~\cite{Altschul2010} and four-fermion mechanisms in which Lorentz- and CPT-breaking structures, including bumblebee-type effective potentials, arise dynamically or through radiative corrections~\cite{Gomes2008,Assuncao2017}. Further developments include tensor bumblebee models~\cite{Potting2024,Assuncao2019}, radiatively induced Lorentz-breaking terms in gauge and gravity sectors~\cite{Ferrari2007,Gomes2010}, and CPT-even or aether-like structures in Lorentz-violating extensions of QED~\cite{Mariz2018,Scarpelli2013}. These works provide useful precedents for the present analysis, where the Lorentz-violating background is not introduced through a fundamental vector field, but arises as a nonzero electromagnetic vacuum configuration selected dynamically by the Hamiltonian structure.

Nonlinear electrodynamics offers a different setting in which vacuum structure, stability, and propagation are naturally intertwined. Beyond Maxwell theory, nonlinear electromagnetic models exhibit a much richer constitutive and dynamical structure. Early work by Boillat clarified general aspects of the dynamics and wave propagation in nonlinear electromagnetic theories \cite{Boillat1970}. Later, it was shown that light propagation in nonlinear electrodynamics can be described in terms of effective geometries \cite{Novello2000}, and that the corresponding Fresnel equation generically leads to birefringence unless additional structural conditions are satisfied \cite{ObukhovRubilar2002}. Nonlinear electrodynamics also provides a natural arena for the study of electric-magnetic duality, as emphasized in the classic work of Gibbons and Rasheed \cite{GibbonsRasheed1995}. These developments show that vacuum structure, propagation, and Hamiltonian consistency are deeply intertwined in nonlinear electrodynamics.

In this context, the Pleba\'nski first-order formulation provides a particularly convenient starting point \cite{Pleb}. In this formalism, the theory is written in terms of an antisymmetric tensor and the gauge potential as independent variables, and the natural Hamiltonian degrees of freedom are the analogues of electric displacement and magnetic field intensity. For theories of the form $\hat{V}=\hat{V}(P)$, this formulation is especially well suited to Dirac's analysis of constraints and leads directly to an effective Hamiltonian description. In previous work \cite{EscobarPotting2020}, a complete Hamiltonian analysis of this class of models was carried out, showing that nontrivial constant electromagnetic backgrounds may arise while preserving gauge invariance, provided certain conditions involving the nonlinear potential are satisfied. In particular, the existence of Lorentz-violating vacua is controlled by a distinguished combination of derivatives of the potential, which also signals a degeneracy in the reduced symplectic structure.

The purpose of the present work is to move beyond that general structural analysis and examine explicit realizations of gauge-invariant nonlinear electrodynamics of the form $\hat{V}(P)$ within the Pleba\'nski Hamiltonian framework. Our aim is to identify concrete models for which nontrivial constant electromagnetic vacua are compatible with an effective Hamiltonian bounded from below and a positive semidefinite Hessian. To this end, we consider three structurally distinct two-parameter nonlinearities---rational asymmetric, logarithmic, and exponential---and determine the regions of parameter space in which these requirements can be simultaneously satisfied.

A central outcome of our analysis is that the coexistence of boundedness, local stability, and spontaneous Lorentz symmetry breaking is highly nontrivial within the class of single-invariant theories. In the explicit models analyzed here, physically admissible and locally stable Lorentz-violating vacua are realized only in the magnetic branch.
By contrast, the electric branch does not yield physically admissible symmetry-breaking vacua, even though stationary configurations may exist locally. We also examine several additional one-parameter models, which show that boundedness of the effective Hamiltonian alone is not sufficient to ensure spontaneous Lorentz symmetry breaking. Finally, we discuss how the symmetry-breaking conditions are related to known strong-field causality criteria in single-invariant nonlinear electrodynamics.

This paper is organized as follows. In Sec.~\ref{Pleb1} we briefly review Pleba\'nski's first-order formulation of nonlinear electrodynamics. In Sec.~\ref{Hamiltonian1} we summarize the Hamiltonian framework and the conditions relevant to vacuum structure and local stability. In Sec.~\ref{Models1} we analyze several explicit $\hat{V}(P)$ models, discuss the obstruction affecting the electric branch, and comment on the relation to causality. Final remarks are presented in Sec.~\ref{Conclu}.


\section{Nonlinear electrodynamics in the Pleba\'nski first-order formulation}
\label{Pleb1}

In this section we briefly summarize the Pleba\'nski first-order formulation of nonlinear electrodynamics, restricting attention to the class of gauge-invariant theories relevant to the present work. Our purpose is to introduce the variables and relations needed for the Hamiltonian analysis developed in the next section.

The Pleba\'nski framework applies more generally to nonlinear electrodynamics depending on both electromagnetic invariants. We therefore begin by recalling the standard description in terms of $\mathcal{F}$ and $\mathcal{G}$, and subsequently restrict attention to the single-invariant class $\mathcal{L}=\mathcal{L}(\mathcal{F})$ relevant to the present work.

In the standard formulation, nonlinear electrodynamics is usually described in terms of a Lagrangian density $\mathcal{L}=\mathcal{L}(\mathcal{F},\mathcal{G})$ depending on the electromagnetic field tensor $F_{\mu\nu}$ through the Lorentz invariants

\begin{equation}
\label{Eq1}
\mathcal{F}=\frac14 F_{\mu\nu}F^{\mu\nu}, \quad \mathcal{G}=\frac14 F_{\mu\nu}\tilde{F}^{\mu\nu},
\end{equation}

In vacuum, this formulation is expressed in terms of the electric and magnetic fields

\begin{equation}
E_i = F_{0i}, \qquad
B_i = \frac12 \epsilon_{ijk}F_{jk},
\end{equation}

where the invariants $\mathcal{F}$ and $\mathcal{G}$  take the form

\begin{equation}
\mathcal{F} = \frac12 (B^2 - E^2), \quad \mathcal{G}=-\vec{E}\cdot\vec{B},
\end{equation}

with $B^2=\vec{B}\cdot\vec{B}$ and  $E^2=\vec{E}\cdot\vec{E}$. While this formulation is natural from the Lagrangian perspective, the Hamiltonian analysis of such theories can be technically nontrivial because the canonical momenta are nonlinear functions of the electric and magnetic fields.

A convenient alternative description was introduced by Pleba\'nski \cite{Pleb}, where the theory is written in first-order form. In this formulation the antisymmetric tensor $P^{\mu\nu}$ and the potential $A_\mu$ are treated as independent variables. The corresponding first-order Lagrangian, without external sources,

\begin{equation}
\hat{\mathcal{L}}(P^{\mu\nu},A_\nu)=-P^{\mu\nu}\partial_\mu A_\nu-\hat{V}(P,Q)
\end{equation}

preserves gauge invariance and allows a systematic application of Dirac's constraint analysis.

An important feature of this formulation is that the natural Hamiltonian variables are not the electric and magnetic fields $(E_i,B_i)$ but instead the pair of fields $(D_i,H_i)$, which are related to the standard variables through the constitutive relations determined by the nonlinear Lagrangian. In analogy with electrodynamics in media, $D_i$ and $H_i$ can be interpreted as the electric displacement and magnetic field intensity.

In the Pleba\'nski framework the invariants

\begin{equation}
\label{Eq5}
P=\frac14 P_{\mu\nu}P^{\mu\nu}, \quad Q= \frac{1}{4}P_{\mu\nu}\tilde{P}^{\mu\nu}
\end{equation}

are constructed from the antisymmetric tensor $P_{\mu\nu}$ that is conjugate to $F_{\mu\nu}$. The components of $P_{\mu\nu}$ are directly related to the fields $D_i$ and $H_i$ through

\begin{equation}
D_i = P_{0i}, \qquad
H_i = -\frac12 \epsilon_{ijk} P_{jk}.
\label{DyH}
\end{equation}

The constitutive relations take the form
\begin{equation}
P^{\mu\nu}=-\mathcal{L}_F F^{\mu\nu}-\mathcal{L}_G \tilde{F}^{\mu\nu},
\qquad
F^{\mu\nu}=-\hat{V}_P P^{\mu\nu}-\hat{V}_Q \tilde{P}^{\mu\nu}.
\label{ERel}
\end{equation}
Here the subscripts in $\mathcal{L}$ and $\hat{V}$ denote derivatives with respect to the corresponding invariant. In terms of $D_i$ and $H_i$ the invariants $P$ and $Q$ can be written as

\begin{equation}
P = \frac12 (H^2 - D^2), \quad Q=-\vec{D}\cdot\vec{H}.
\end{equation}

The Pleba\'nski first-order formulation is particularly useful because it rewrites nonlinear electrodynamics in terms of variables naturally adapted to the Hamiltonian description. Instead of treating the electric field as the variable conjugate to the vector potential, one introduces the antisymmetric tensor \(P^{\mu\nu}\), whose spatial components are identified with the electric displacement \(D_i\) and the magnetic field intensity \(H_i\).

In this way, the nonlinearity of the theory is encoded in the constitutive relation between \(F_{\mu\nu}\) and \(P^{\mu\nu}\), or equivalently in the potential \(\hat V(P,Q)\), while gauge invariance remains manifest. This formulation also makes transparent the analogy with electrodynamics in media and provides a natural setting for Dirac's analysis of constraints, as well as for the study of fluctuations on nontrivial electromagnetic backgrounds. For the present analysis, its main advantage is that it leads directly to the effective Hamiltonian and to a characterization of nontrivial vacua compatible with the constraint structure.

We now specialize to the single-invariant sector \(\mathcal{L}=\mathcal{L}(\mathcal{F})\), where the nonlinear dynamics is encoded in a potential \(\hat V(P)\).

\subsection{Relation between the two formulations}
\label{Rel1}
The standard and first-order descriptions of nonlinear electrodynamics are related by a Legendre-type transformation. 
In the usual formulation, the theory is specified by a Lagrangian density $\mathcal{L}=\mathcal{L}(\mathcal{F},\mathcal{G})$, whereas in the Pleba\'nski formulation the nonlinear dynamics is encoded in the potential $\hat V=\hat V(P,Q)$.

The two descriptions are connected through the constitutive relations in Eq. (\ref{ERel}). The potential \(\hat V(P,Q)\) is obtained from \(\mathcal{L}(\mathcal{F},\mathcal{G})\) through
\[
\hat V(P,Q)
=
-\mathcal{L}(\mathcal{F},\mathcal{G})
+
2\left(
\mathcal{F}\mathcal{L}_{\mathcal F}
+
\mathcal{G}\mathcal{L}_{\mathcal G}
\right),
\]
where, after the transformation, the invariants \(\mathcal{F}\) and \(\mathcal{G}\) must be expressed in terms of \(P\) and \(Q\). Equivalently, starting from a given Pleba\'nski potential, the corresponding Lagrangian is recovered as
\[
\mathcal{L}(\mathcal{F},\mathcal{G})
=
-\hat V(P,Q)
+
2\left(
P\hat V_P
+
Q\hat V_Q
\right),
\]
with \(P\) and \(Q\) understood as functions of \(\mathcal{F}\) and \(\mathcal{G}\), using the same conventions for the dual tensor as those used in Eqs.~(\ref{Eq1})-(\ref{Eq5}).


A useful way to make this correspondence explicit is to consider a nonlinear model for which both formulations are known in closed form. The Born--Infeld theory provides such an example. In the usual second-order formulation the Born--Infeld action is
\[
S_{\rm BI}[A_\mu]
=
\int d^4x\,
\mathcal{L}_{\rm BI}(\mathcal{F},\mathcal{G}),
\]
with the Lagrange density
\[
\mathcal{L}_{\rm BI}(\mathcal{F},\mathcal{G})
=
b^2
\left[
1-
\sqrt{
1+\frac{2\mathcal{F}}{b^2}
-\frac{\mathcal{G}^2}{b^4}
}
\right],
\]

where \(b\) is the Born--Infeld parameter. Defining
\[
R
=
\sqrt{
1+\frac{2\mathcal{F}}{b^2}
-\frac{\mathcal{G}^2}{b^4}
},
\]
one has
\[
\mathcal{L}_{\mathcal F}
=
-\frac{1}{R},
\qquad
\mathcal{L}_{\mathcal G}
=
\frac{\mathcal{G}}{b^2R}.
\]
The constitutive relation in Eq. (\ref{ERel}), 
$P^{\mu\nu}
=
-\mathcal{L}_{\mathcal F}F^{\mu\nu}
-
\mathcal{L}_{\mathcal G}\widetilde F^{\mu\nu}$,
 gives
\[
P^{\mu\nu}
=
\frac{1}{R}
\left(
F^{\mu\nu}
-
\frac{\mathcal{G}}{b^2}\widetilde F^{\mu\nu}
\right).
\]
Contracting this relation with itself and with its dual, and using
\[
\frac14 F_{\mu\nu}F^{\mu\nu}=\mathcal{F},
\qquad
\frac14 F_{\mu\nu}\widetilde F^{\mu\nu}=\mathcal{G},
\qquad
\frac14 \widetilde F_{\mu\nu}\widetilde F^{\mu\nu}=-\mathcal{F},
\]
one obtains
\[
P
=
\frac{
\left(1-\frac{\mathcal{G}^2}{b^4}\right)\mathcal{F}
-
\frac{2\mathcal{G}^2}{b^2}
}{
1+\frac{2\mathcal{F}}{b^2}
-\frac{\mathcal{G}^2}{b^4}
},
\qquad
Q=\mathcal{G}.
\]
Equivalently,
\[
\mathcal{G}=Q,
\qquad
\mathcal{F}
=
\frac{
P+\frac{2Q^2}{b^2}
-\frac{P Q^2}{b^4}
}{
1-\frac{2P}{b^2}
-\frac{Q^2}{b^4}
}.
\]
Substituting these relations into the Legendre transform
\[
\hat V(P,Q)
=
-\mathcal{L}(\mathcal{F},\mathcal{G})
+
2\left(
\mathcal{F}\mathcal{L}_{\mathcal F}
+
\mathcal{G}\mathcal{L}_{\mathcal G}
\right),
\]
one obtains, after simplification,
\[
\hat V_{\rm BI}(P,Q)
=
b^2
\left[
\sqrt{
1-\frac{2P}{b^2}
-\frac{Q^2}{b^4}
}
-1
\right].
\]
Thus the equivalent first-order Pleba\'nski action is
\[
\hat S_{\rm BI}[A_\mu,P_{\mu\nu}]
=
\int d^4x\,
\left[
-P^{\mu\nu}\partial_\mu A_\nu
-
\hat V_{\rm BI}(P,Q)
\right].
\]

Indeed, the derivatives of this potential are
\[
\hat V_P
=
-\frac{1}{
\sqrt{
1-\frac{2P}{b^2}
-\frac{Q^2}{b^4}
}
},
\qquad
\hat V_Q
=
-\frac{Q/b^2}{
\sqrt{
1-\frac{2P}{b^2}
-\frac{Q^2}{b^4}
}
},
\]
so that the inverse constitutive relation takes the expected Pleba\'nski form
\[
F^{\mu\nu}
=
-\hat V_P P^{\mu\nu}
-
\hat V_Q \widetilde P^{\mu\nu}.
\]
This example shows explicitly how a nonlinear electrodynamics model written in the usual variables \((\mathcal{F},\mathcal{G})\) is equivalently encoded, on a regular branch of the constitutive map, in a first-order Pleba\'nski potential \(\hat V(P,Q)\).

In the present work we restrict attention to the single-invariant sector. Thus, the standard formulation is described by $\mathcal{L}=\mathcal{L}(\mathcal{F})$,
and the Pleba\'nski formulation by $\hat V=\hat V(P)$. In this case the constitutive relations (\ref{ERel}) reduce to
\[
P^{\mu\nu}
=
-\mathcal{L}_{\mathcal F}F^{\mu\nu},
\qquad
F^{\mu\nu}
=
-\hat V_P P^{\mu\nu}.
\]
Consequently, the two invariants are related by
\[
P
=
\mathcal{F}\,\mathcal{L}_{\mathcal F}^{\,2},
\qquad
\mathcal{F}
=
P\,\hat V_P^{\,2}.
\]

These relations are understood on regular branches of the constitutive map, where the transformation between the two sets of invariants is locally invertible $\mathcal{F}\leftrightarrow P$. In the single-invariant sector this requires
\[
\frac{d\mathcal{F}}{dP}
=
\hat V_P\left(\hat V_P+2P\hat V_{PP}\right)
\neq 0.
\]
A Pleba\'nski potential \(\hat V(P)\) therefore determines the corresponding standard Lagrangian $\mathcal{L}(\mathcal{F})$ only after solving, at least locally, the implicit relation
\[
\mathcal{F}=P\hat V_P^{\,2}
\]
for \(P=P(\mathcal{F})\). For the explicit models analyzed in this work, this inversion cannot be carried out in closed elementary form. Consequently, the associated standard formulation \(\mathcal{L}(\mathcal{F})\) is most naturally specified parametrically as
\begin{equation}
\mathcal{F}(P)=P\hat V_P^{\,2},
\qquad
\mathcal{L}(P)=-\hat V(P)+2P\hat V_P(P).
\end{equation}
This does not obstruct the definition of the theory. Rather, it reflects the fact that the Pleba\'nski formulation provides the natural description of the models considered here, particularly near the degenerate vacuum configurations relevant for the spontaneous Lorentz symmetry breaking analysis. Indeed, as will be seen below, the symmetry-breaking conditions are precisely associated with the vanishing of one of the factors in \(d\mathcal F/dP\). Thus, in the following we take the Pleba\'nski potential \(\hat V(P)\) as the primary object defining the model, while the associated standard Lagrangian \(\mathcal{L}(\mathcal{F})\), when needed, is understood parametrically on regular branches away from such degeneracies.

\section{Hamiltonian structure}
\label{Hamiltonian1}

In this section we summarize the Hamiltonian framework for the single-invariant sector introduced above, following the Pleba\'nski formulation and the analysis of Ref.~\cite{EscobarPotting2020}. Our purpose is to establish the notation and the conditions relevant to vacuum structure and local stability in the models considered below.

Starting from the first-order Lagrangian where $P^{\mu\nu}$ and $A_\mu$ are treated as independent variables, the canonical momenta lead to a set of primary constraints. Implementing Dirac's procedure for constrained systems \cite{Dirac1}, one finds that the theory possesses both first-class and second-class constraints.

The first-class constraints generate the gauge symmetry of the theory, while the second-class constraints, together with the effective Hamiltonian, determine the dynamics through Dirac brackets.

 The structure of these  second-class constraints can be characterized by the matrix of Poisson brackets among them. As shown in Ref. \cite{EscobarPotting2020}, the determinant of this matrix is proportional to a quantity that depends on the derivatives of the nonlinear potential, explicitly 

\begin{equation}
S = H^2 \hat{V}_{PP} + \hat{V}_P,
\label{C1}
\end{equation}

When $S\neq 0$ the matrix of Poisson brackets is non-singular and the constraint structure is regular. However, when $S=0$ the determinant vanishes and the constraint matrix becomes degenerate. As already anticipated in the introduction, this degeneracy also plays a central role in the existence of nontrivial vacuum configurations.

After implementing the constraints and eliminating the unphysical variables, the dynamics can be expressed in terms of the three components of the fields $D_i$ and $H_i$. The resulting effective Hamiltonian density, which arises from the canonical Hamiltonian after the second-class constraints have been strongly imposed, takes the form

\begin{equation}
\mathcal{H}_{\mathrm{eff}}(D_i,H_i)
=- H^2 \hat{V}_P + \hat{V}(P).
\label{HEffec}
\end{equation}

\subsection{Stationary configurations}
\label{Hessian1}

It should be emphasized that,  in the present constrained framework, vacuum configurations are obtained by extremizing the effective Hamiltonian rather than the potential itself, in contrast with the standard situation in unconstrained field theories. The stationarity conditions are obtained by taking derivatives with respect to the canonical variables,

\begin{equation}
\frac{\partial \mathcal{H}_{\mathrm{eff}}}{\partial D_i}=0,
\qquad
\frac{\partial \mathcal{H}_{\mathrm{eff}}}{\partial H_i}=0 .
\end{equation}

Using the explicit form of $\mathcal{H}_{\mathrm{eff}}$ these conditions reduce to

\begin{align}
\label{C12}
D_{i} ( H^2 \hat{V}_{PP} - \hat{V}_P ) &=0 , \\
H_{i} ( H^2 \hat{V}_{PP} + \hat{V}_P ) &=0 .
\label{C13}
\end{align}

We consider two branches of stationary configurations, namely: a) $D_i=0$ and $H_i\neq0$ (magnetic vacua) and b) $D_i\neq0$ and $H_i=0$ (electric vacua). For the nontrivial magnetic branch, the stationarity condition reduces to

\begin{equation}
 S_m=H^2 \hat{V}_{PP} + \hat{V}_P =0.
 \label{SM}
\end{equation}

 The subscript \(m\) indicates that this expression corresponds to the magnetic branch, namely the form taken by the general stationarity condition once the magnetic vacuum conditions have been applied. Similarly, the notation \(S_e\) will be used for the electric vacuum branch.

The condition in Eq.~(\ref{SM}), \(S_m=0\), defines the hypersurface in field space where spontaneous Lorentz symmetry breaking can occur. It is the magnetic-branch specialization of the general condition in Eq.~(\ref{C1}), and therefore identifies the locus where the constraint structure becomes degenerate. However, not every configuration satisfying \(S_m=0\) corresponds to a physically admissible spontaneously broken vacuum. One must further require that the associated stationary point does not correspond to a local maximum. The analysis is therefore reduced to studying the eigenvalues of the Hessian matrix of the effective Hamiltonian. For the nontrivial electric vacuum branch, a similar conclusion follows: the stationarity condition reduces to \(\hat{V}_P=0\), and hence \(S_e=\hat{V}_P\).

\subsection{Hessian matrix and local stability}
\label{H12}

The local stability of a stationary configuration is determined by the Hessian matrix of the effective Hamiltonian with respect to the canonical variables,

\begin{equation}
\mathcal{M}_{ab}
=
\frac{\partial^2 \mathcal{H}_{\mathrm{eff}}}
{\partial X_a \partial X_b},
\qquad
X_a = (D_i,H_i).
\end{equation}

Evaluating these derivatives yields the Hessian matrix

\begin{equation}
\mathcal{M}_{ab}
=\left(
\begin{array}{cc}
2\mathcal{H}_{D^2}\,\delta_{ij}
+4\mathcal{H}_{D^2D^2}\,D_iD_j
&
4\mathcal{H}_{D^2H^2}D_iH_j
\\[1.2ex]

4\mathcal{H}_{D^2H^2}H_iD_j
&
2\mathcal{H}_{H^2}\,\delta_{ij}
+4\mathcal{H}_{H^2H^2}\,H_iH_j
\end{array}
\right),
\qquad i,j=1,2,3,
\end{equation}

where, for simplicity, we have omitted the subscript ``$\mathrm{eff}$'' in the Hamiltonian and the subscripts denote the derivative with respect to the quantities $H^2$ or $D^2$, explicitly

\begin{eqnarray}
&\mathcal{H}_{H^2}=-\dfrac{1}{2}\left(\hat{V}_P+H^2\hat{V}_{PP}\right),\quad
\mathcal{H}_{D^2}=\dfrac{1}{2}\left(H^2\hat{V}_{PP}-\hat{V}_P\right),\quad
\mathcal{H}_{H^2H^2}=-\dfrac{1}{4}\left(3\hat{V}_{PP}+H^2\hat{V}_{PPP}\right),\\[6pt]
&\mathcal{H}_{D^2D^2}=\dfrac{1}{4}\left(\hat{V}_{PP}-H^2\hat{V}_{PPP}\right),\quad
\mathcal{H}_{D^2H^2}=\dfrac{1}{4}\left(\hat{V}_{PP}+H^2\hat{V}_{PPP}\right).
\end{eqnarray}

The eigenvalues of the matrix $\mathcal{M}_{ab}$ determine the spectrum of quadratic fluctuations around the vacuum configuration. A stationary point will be energetically stable provided the Hessian is positive semidefinite and the effective Hamiltonian is bounded from below. This requirement is in line with the general lesson from Lorentz-breaking field theories: the existence of a nonzero vacuum configuration is not by itself sufficient, and Hamiltonian boundedness is needed to avoid unstable branches~\cite{Potting2024}.

As shown in Ref.~\cite{EscobarPotting2020}, when the vacuum satisfies the condition $S=0$ the Hessian generically develops zero eigenvalues associated with the spontaneously broken Lorentz symmetry. The remaining eigenvalues depend on the specific form of the nonlinear potential and must be analyzed on a model-by-model basis.

In the following, for simplicity, we introduce the variables
\begin{equation}
h:=\vec H\cdot \vec H, \qquad d:=\vec D\cdot \vec D,
\label{hd2}
\end{equation}
and we may write
\begin{equation}
P=\frac{1}{2}(h-d),
\qquad h\ge 0,\qquad d\ge 0.
\end{equation}

In terms of these variables, at the stationary configuration defined by $D_i=0$ and $H_i\neq0$ (magnetic vacua), the Hessian matrix takes the block-diagonal form
\begin{equation}
\mathcal{M}_{ab,m}
=
\begin{pmatrix}
2\mathcal{H}_{d}\,\delta_{ij} & 0_{ij}\\[1ex]
0_{ij} & 4\mathcal{H}_{hh}\,H_iH_j
\end{pmatrix},
\qquad i,j=1,2,3.
\label{eq:rational_hessian_vacuum}
\end{equation}

Its eigenvalue structure can be read off directly. The block proportional to $\delta_{ij}$ contributes a triple eigenvalue, $\mu_{1}=\mu_{2}=\mu_{3}=2\mathcal{H}_{d}$,
whereas the submatrix $4\mathcal{H}_{hh}H_iH_j$ has rank one and therefore contributes a single nonvanishing eigenvalue $\mu_{4}=4h\,\mathcal{H}_{hh}$, together with two vanishing eigenvalues. Therefore, the Hessian is positive semidefinite provided $\mathcal{H}_{d}>0$ and $\mathcal{H}_{hh}>0$. In that case the stationary point possesses four strictly positive eigenvalues and two zero modes. The latter are naturally interpreted as flat directions associated with the spontaneous breaking of spatial rotational symmetry by the vacuum configuration $ D_i=0$, $H_i\neq 0$.

In the present case, the existence of a physically admissible Lorentz-violating vacuum requires more than the mere existence of a stationary point. First, the vacuum configuration must satisfy the condition \(S_m=h\hat{V}_{PP}+\hat{V}_P=0\), which determines the vacuum value \(h=h_0\). Since \(h_0=\vec{H}_0\cdot\vec{H}_0\) is a squared magnitude, it must be real and nonnegative. Second, local stability requires that the Hessian matrix of the effective Hamiltonian be positive semidefinite. Therefore, spontaneous Lorentz symmetry breaking with a stable magnetic vacuum occurs only when the following requirements are simultaneously satisfied:
\begin{equation}
S_m=0,\qquad h_0>0,\qquad \mathcal{H}_d>0,\qquad \mathcal{H}_{hh}>0.
\label{CR2}
\end{equation}
These are the conditions that will be used in Sec.~\ref{Models1} to identify physically admissible magnetic vacua in explicit models.

It is important to emphasize that the spontaneous Lorentz symmetry breaking considered here differs conceptually from the usual bumblebee mechanism. In the latter, a vector field acquires a nonzero vacuum expectation value through an explicit symmetry-breaking potential, for example $V(A_\mu A^\mu\pm b^2)\Rightarrow \langle A_\mu A^\mu\rangle=\mp b^2$. In the present gauge-invariant framework, by contrast, the action depends only on derivatives of the gauge potential \(A_\mu\), and the gauge potential itself does not acquire a vacuum expectation value. The Lorentz-violating vacuum is instead associated with a nontrivial stationary configuration of the reduced effective Hamiltonian, $\mathcal{H}_{\rm eff}$ given in Eq. (\ref{HEffec}), in terms of the first-order fields \(D_i\) and \(H_i\).

More precisely, the stationary conditions obtained by extremizing \(\mathcal{H}_{\rm eff}\) admit a magnetic branch
\[
D_i=0,
\qquad
H_i=H_{0i}\neq0 .
\]

where \(H_{0i}\) is a constant background field. Thus, the role usually played by the symmetry-breaking potential in bumblebee models is played here by the effective Hamiltonian \(\mathcal H_{\rm eff}\), while the vacuum configuration is the constant magnetic background \(H_{0i}\), which corresponds to the spatial part of \(P_{\mu\nu}^{(0)}\). This background selects a preferred spacetime direction and therefore breaks the Lorentz symmetry. Section \ref{Models1} is devoted to showing explicit examples of this mechanism.

For electric vacuum configurations, \(D_i\neq 0\) and \(H_i=0\), the stationarity condition implies \(\hat V_{P}=0\). Under these conditions, the Hessian matrix reduces to
\begin{equation}
\mathcal{M}_{ab,e}=
\begin{pmatrix}
 4\mathcal{H}_{dd}\,D_iD_j & 0_{ij}\\[1ex]
0_{ij} & 0_{ij}
\end{pmatrix},
\qquad i,j=1,2,3.
\end{equation}
Its eigenvalue structure can again be read off directly. Since the matrix \(D_iD_j\) has rank one, the electric block contributes a single nonvanishing eigenvalue,
\begin{equation}
\mu_{1}= 4d\mathcal{H}_{dd}=\hat V_{PP}\,d,
\qquad d=\vec D\cdot\vec D,
\end{equation}
while the remaining five eigenvalues vanish,
\begin{equation}
\mu_{2}=\mu_{3}=\mu_{4}=\mu_{5}=\mu_{6}=0.
\end{equation}
Therefore, the Hessian is highly degenerate in the electric branch: at quadratic order, curvature is present only along the direction parallel to \(\vec D\), whereas all transverse electric directions and the whole magnetic sector remain flat.

This structure has a natural symmetry interpretation. A nontrivial electric background \(\vec D\neq 0\) selects a preferred spatial direction and therefore breaks spatial rotations down to the subgroup that leaves \(\vec D\) invariant. Accordingly, two of the zero modes may be interpreted as the expected flat directions associated with infinitesimal rotations of the vacuum orientation. The additional zero modes, however, indicate a stronger degeneracy of the quadratic fluctuation operator, showing that the electric branch is not fully stabilized at Hessian level and that its analysis requires special care beyond the quadratic approximation.

At the level of branchwise stationarity and Hessian positivity, the electric branch is characterized by
\begin{equation}
S_e=\hat{V}_P=0,\qquad d_0>0,\qquad \mathcal{H}_{dd}>0.
\label{CR3}
\end{equation}

In the following section we apply this framework to explicit nonlinear electromagnetic models in order to determine their vacuum structure, local stability, and the compatibility of these properties with spontaneous Lorentz symmetry breaking.

\section{Explicit $\hat{V}(P)$ Models: Vacuum Structure, Stability, and Causality}
\label{Models1}

\subsection{Rational asymmetric model}

As a first explicit example, we consider the rational asymmetric nonlinear electrodynamics model
\begin{equation}
\hat V(P)= -P+\frac{\lambda P^{3}}{1+\eta P^{2}},
\label{eq:rational_VP}
\end{equation}

where $\lambda$ and $\eta$ are real parameters. Equivalently, in the first-order formulation with the variables \(D_i\) and \(H_i\) defined in Eq. (\ref{DyH}), and using \(P=\frac12(H^2-D^2)\), the same potential takes the form
\[
\hat V(D,H)
=
-\frac12(H^2-D^2)
+
\frac{\lambda}{8}
\Bigg[\frac{(H^2-D^2)^3}
{1+\frac{\eta}{4}(H^2-D^2)^2}\bigg].
\]
This form makes explicit that \(P\) is not an additional independent quantity, but the scalar invariant constructed from the Pleba\'nski first-order variables. This model is particularly interesting because it provides a non-polynomial deformation of Maxwell electrodynamics while remaining within the class of theories depending only on the invariant $P$. The corresponding standard Lagrangian, in terms of $\mathcal{F}$, is obtained through the Legendre map of Sec.~\ref{Rel1}. It is most naturally written in parametric form as
\[
\mathcal{L}(P)
=
-P+
\lambda P^3
\frac{5+\eta P^2}{(1+\eta P^2)^2},
\qquad
\mathcal{F}(P)
=
P\left[
-1+\lambda
\frac{P^2(3+\eta P^2)}{(1+\eta P^2)^2}
\right]^2 .
\]
Thus, Eq.~(\ref{eq:rational_VP}) defines, at least locally on a chosen Legendre branch, an \(\mathcal{L}(\mathcal{F})\) theory, with \(P=P(\mathcal{F})\) obtained implicitly from the second relation.

Given the potential in Eq. (\ref{eq:rational_VP}), from Eq. (\ref{HEffec}) with $h=H^{2}$, after straightforward substitution we can derive

\begin{equation}
\mathcal H_{\rm eff}(h,d)=
\frac{d-h}{2}
+\frac{\lambda (h-d)^{3}}{8\left(1+\frac{\eta}{4}(h-d)^{2}\right)}
+
\frac{h\left[1+\frac{1}{4}(2\eta-3\lambda)(d-h)^{2}
+\frac{1}{16}\eta(\eta-\lambda)(d-h)^{4}\right]}
{\left(1+\frac{\eta}{4}(d-h)^{2}\right)^{2}}.
\label{eq:rational_Heff_hd}
\end{equation}

To determine whether the effective Hamiltonian of the rational asymmetric model is bounded from below, one must analyze its behavior along the asymptotic directions of the physical domain defined by $h \geq 0$, $d \geq 0$.

Rather than attempting a direct global analysis over the full $(h,d)$ plane, it is sufficient to examine the limits in which one or both variables become arbitrarily large, since any possible instability would necessarily appear along one of these directions.

The first relevant limit is $h\to\infty$ with $d$ kept fixed. This probes the behavior of the effective Hamiltonian when the magnetic sector dominates. The second limit is $d\to\infty$ with $h$ fixed, which tests the opposite situation, namely whether the electric sector can drive the Hamiltonian to arbitrarily negative values. Finally, one must also consider the simultaneous limit $h,d\to\infty$ while keeping the difference $h-d$ constant. This direction turns out to be particularly important, since it explores configurations in which both invariants grow without bound while their difference remains finite, and it provides the strongest restriction on the parameter space.

The outcome of this asymptotic analysis is that the effective Hamiltonian is strictly bounded from below provided the parameters satisfy

\begin{equation}
\eta>0,
\qquad
\eta \geq \frac{9}{8}\lambda.
\label{etalamb}
\end{equation}

These inequalities define the region of parameter space for which the rational asymmetric model does not develop runaway directions toward arbitrarily negative energy. Therefore, within this domain, the model provides a physically admissible candidate for the realization of nontrivial Lorentz-violating vacua.

Once this region is identified, one can proceed to the analysis of nontrivial stationary configurations satisfying the vacuum condition, and to the corresponding Hessian matrix in order to determine whether the resulting Lorentz-violating vacuum is locally stable. At this point, we focus on the magnetic case, defined by configurations $D_{i}=0$ and $H_{i}\neq0$, or equivalently by $d=0$ and $h\neq0$.

For the rational asymmetric model, the vacuum value $h_0$ is fixed by the stationarity condition $S_m=0$. In terms of \(h\), this condition reduces to the algebraic equation
\begin{equation}
(\eta^{3}-\eta^{2}\lambda)\,h^{6}
+12\eta^{2}h^{4}
+(48\eta-240\lambda)h^{2}
+64=0.
\label{eq:rational_h0_polynomial}
\end{equation}
Therefore, \(h_0\) is determined as a real positive root of Eq.~\eqref{eq:rational_h0_polynomial}. Since \(h_0=\vec H_{0}\cdot \vec H_{0}\) represents the squared magnitude of the vacuum magnetic background, only the positive real solution is physically admissible.

 For the rational asymmetric model, the conditions in Eq. (\ref{CR2}) restrict the parameter space to
\begin{equation}
\frac{9}{8}\lambda \leq \eta < \left(\frac{25}{16}+\frac{5\sqrt{5}}{16}\right)\lambda,
\label{eq:rational_ssb_region}
\end{equation}

This interval defines the region in parameter space where the model admits nontrivial Lorentz-violating magnetic vacua with a positive semidefinite Hessian and therefore with local stability.

A detailed inspection of the parameter space shows that, for the case \(H_i=0\) and \(D_i\neq 0\) (electric branch), the branchwise stationarity and Hessian conditions in Eq.~(\ref{CR3}) are incompatible with the condition that the effective Hamiltonian be bounded from below. Hence, for the rational asymmetric model, there is no region in the \((\lambda,\eta)\) plane in which an electric branch satisfying Eq.~(\ref{CR3}) coexists with global energetic admissibility.

If the requirement that the effective Hamiltonian be globally bounded from below is relaxed, the electric branch may still satisfy Eq.~(\ref{CR3}) in sectors excluded by the boundedness condition. Explicitly, this occurs when
\[
\lambda<\eta<\frac{9}{8}\lambda,
\]
which is incompatible with the boundedness condition in Eq. (\ref{etalamb}). The magnitude of \(d_0\) is given by the positive real root of the following algebraic equation

\begin{equation}
   d^4 \left(\eta ^2-\eta  \lambda \right)+d^2 (8 \eta -12 \lambda )+16=0.
   \label{ElectricR}
\end{equation} 

However, the general analysis of the electric branch given below shows that such configurations are destabilized by arbitrarily small magnetic perturbations, so they cannot represent local minima of the full effective Hamiltonian.

Thus, in the rational asymmetric model, physically admissible Lorentz-violating vacua arise only in the magnetic branch.

\subsection{Logarithmic model}

As a second explicit example, we consider a logarithmic nonlinear electrodynamics model of the form

\begin{equation}
\hat{V}(P)= -P - \frac{\lambda}{\eta}\ln(1+\eta P).
\label{eq:log_VP}
\end{equation}

where $\lambda$ and $\eta$ are real parameters. In terms of the first-order variables this becomes
\[
\hat V(D,H)
=
-\frac12(H^2-D^2)
-
\frac{\lambda}{\eta}
\ln\!\left[
1+\frac{\eta}{2}(H^2-D^2)
\right].
\]
Thus the logarithmic nonlinearity depends on the same Pleba\'nski invariant
\(P=\frac12(H^2-D^2)\), expressed directly in terms of the Hamiltonian fields ($D_i,H_i$).

This model provides a qualitatively different deformation of Maxwell electrodynamics from the rational case, while still remaining within the class of gauge-invariant theories depending only on the invariant $P$. 
Using the Legendre construction summarized in Sec.~\ref{Rel1}, the potential in Eq.~(\ref{eq:log_VP}) can also be associated with a standard single-invariant Lagrangian. The map is conveniently expressed as
\[
\mathcal{L}(P)
=
-P
-\frac{2\lambda P}{1+\eta P}
+
\frac{\lambda}{\eta}\ln(1+\eta P),\quad
\mathcal{F}(P)
=
P\left(
1+\frac{\lambda}{1+\eta P}
\right)^2.
\]
This parametric representation specifies the corresponding $\mathcal{L}(\mathcal{F})$ model once the second relation is inverted locally.

As in the previous model, in terms of the variables $h$ and $d$ in Eq. (\ref{hd2}), the effective Hamiltonian can be written as follows

\begin{equation}
\mathcal H_{\rm eff}(h,d)
=
\frac{h+d}{2}
+\frac{\lambda h}{1+\frac{\eta}{2}(h-d)}
-\frac{\lambda}{\eta}\ln\!\left(1+\frac{\eta}{2}(h-d)\right).
\label{eq:log_heff_hd}
\end{equation}

Notice that the logarithmic structure imposes the nontrivial domain condition

\begin{equation}
1+\eta P>0,
\qquad\text{or equivalently}\qquad
1+\frac{\eta}{2}(h-d)>0,
\label{eq:log_domain}
\end{equation}
which plays an essential role in the global stability analysis.

The logarithmic model is particularly interesting because the admissible region in field space is no longer the full quadrant $h\geq0$, $d\geq0$, but is restricted by Eq.~\eqref{eq:log_domain}.  A direct asymptotic analysis shows that $\mathcal{H}_{\mathrm{eff}}$ remains bounded from below provided
\begin{equation}
\lambda \ge 0,
\qquad
\eta>0.
\label{LogBound}
\end{equation}
Indeed, in this parameter region the Hamiltonian grows to $+\infty$ along all admissible asymptotic directions in the physical domain, including the approach to the boundary imposed by the logarithmic term. By contrast, for $\lambda<0$ the logarithmic contribution becomes negatively divergent near the boundary, so the Hamiltonian is no longer bounded from below.

We now apply the general criteria established in Sec. \ref{H12}. In particular, a physically admissible Lorentz-violating magnetic vacuum must satisfy the conditions summarized in Eq. (\ref{CR2}), namely the stationarity condition \(S_m=0\), the positivity of the vacuum magnitude \(h_0>0\), and the positivity of the nonvanishing eigenvalues of the Hessian. For the logarithmic model, a direct analysis of these requirements shows that they are simultaneously satisfied in the parameter region
\begin{equation}
\lambda > 8, \qquad \eta > 0.
\label{eq:log_region}
\end{equation}

Within this domain, the stationarity condition \(S_m=0\) fixes the magnitude of the vacuum field to

\begin{equation}
h_{0}
=
\frac{1}{\eta}
\left(
\sqrt{\lambda^{2}-8\lambda}+\lambda-2
\right).
\label{eq:log_h0}
\end{equation}
Therefore, for \(\lambda>8\) and \(\eta>0\), the logarithmic model admits nontrivial Lorentz-violating magnetic vacua with positive vacuum magnitude and a positive semidefinite Hessian, and hence provides a consistent realization of spontaneous Lorentz symmetry breaking.

For the electric branch, \(H_i=0\) and \(D_i\neq 0\), so that \(P=-d/2\), and the stationarity condition reduces to \(\hat V_P=0\). For the logarithmic potential, this implies
\[
d_0=\frac{2(\lambda+1)}{\eta}.
\]
At this point the logarithmic model differs crucially from the other examples, because the argument of the logarithm must satisfy
\[
1+\eta P>0 \quad \Rightarrow \quad 1-\frac{\eta d_0}{2}>0.
\]
Evaluating this condition at the stationary point yields \(-\lambda>0\), and hence necessarily \(\lambda<0\). On the other hand, the electric Hessian has a single nonvanishing eigenvalue \(\mu_1=4d_0\mathcal{H}_{dd}=\hat V_{PP}d_0\), so branchwise Hessian positivity requires \(d_0>0\) and \(\mathcal{H}_{dd}>0\). For the logarithmic model this implies \(\eta<0\), and then \(d_0>0\) further requires \(\lambda<-1\). Thus, at the level of branchwise stationarity and Hessian positivity, the electric branch is realized only in the parameter region

\[
\lambda<-1,\qquad \eta<0.
\]

This result is incompatible with the global boundedness condition in Eq.~(\ref{LogBound}), which requires \(\lambda\ge 0\) and \(\eta>0\). The electric branch is therefore ruled out both by its incompatibility with global boundedness and by the general instability mechanism discussed below. The logarithmic model consequently supports physically admissible Lorentz-violating vacua only in the magnetic branch.

\subsection{Exponential model}

As a third example, we consider an exponential deformation of Maxwell electrodynamics,

\begin{equation}
\hat{V}(P)=-P-\frac{\lambda}{\eta}\left(1-e^{-\eta P}\right).
\label{eq:exp_VP}
\end{equation}

where $\lambda$ and $\eta$ are real parameters, with $\eta\neq 0$. Likewise, in terms of \(D_i\) and \(H_i\), this potential reads
\[
\hat V(D,H)
=
-\frac12(H^2-D^2)
-
\frac{\lambda}{\eta}
\left[
1-
\exp\!\left(
-\frac{\eta}{2}(H^2-D^2)
\right)
\right].
\] In contrast with the logarithmic model, the present nonlinearity does not introduce an additional restriction on the physical domain. This makes the global boundedness analysis more direct, while still allowing for nontrivial Lorentz-violating vacua.

For completeness, the standard single-invariant formulation associated with Eq.~(\ref{eq:exp_VP}) may again be obtained through the Legendre map of Sec.~\ref{Rel1}. In this case one obtains the parametric relations
\[
\mathcal{L}(P)
=
-P
+
\frac{\lambda}{\eta}\left(1-e^{-\eta P}\right)
-
2\lambda P e^{-\eta P},\quad
\mathcal{F}(P)
=
P\left(1+\lambda e^{-\eta P}\right)^2.
\]
Thus, the corresponding $\mathcal{L}(\mathcal{F})$ theory is defined locally by eliminating $P$ between these two expressions.

Using the expression in Eq. (\ref{HEffec}) for the effective Hamiltonian one obtains

\begin{equation}
\mathcal{H}_{\mathrm{eff}}(h,d)
=
\frac{h+d}{2}
-\frac{\lambda}{\eta}
+\lambda\left(h+\frac{1}{\eta}\right)e^{\frac{\eta}{2}(d-h)}.
\label{eq:exp_Heff}
\end{equation}

As in the previous models, the first step is to determine whether the effective Hamiltonian is bounded from below in the physical region $h\geq 0$, $d\geq 0$. In the present case this analysis is particularly transparent, since the potentially dangerous contribution is entirely controlled by the exponential term. For the nontrivial exponential deformation, one finds that the Hamiltonian is bounded from below provided
\begin{equation}
\lambda>0,
\qquad
\eta>0.
\label{eq:exp_bounded}
\end{equation}
The limiting case \(\lambda=0\) reduces to the linear Maxwell theory and will not be considered as a genuinely nonlinear deformation. Under the conditions (\ref{eq:exp_bounded}), no runaway direction toward arbitrarily negative energy exists in the $(h,d)$ plane.

The boundedness result derived above is only the first step in identifying physically admissible vacua. To obtain spontaneous Lorentz symmetry breaking, the exponential model must satisfy additional requirements. 

For the magnetic branch, the conditions in Eq.~(\ref{CR2}) select the parameter region

\begin{equation}
\eta>0,
\qquad
\lambda>\frac{e^{3/2}}{2}.
\label{eq:exp_region}
\end{equation}
In this domain, the vacuum value \(h_0\) is determined by the stationarity condition \(S_m=0\), which yields

\begin{equation}
h_0=\frac{1-2\,W_{-1}\!\left(-\dfrac{e^{1/2}}{2\lambda}\right)}{\eta},
\end{equation}
where \(W_{-1}\) is the \(-1\) branch of the Lambert \(W\) function.

Therefore, for \(\eta>0\) and \(\lambda>e^{3/2}/2\), the exponential model provides a third explicit realization of a nontrivial magnetic Lorentz-violating vacuum with positive vacuum magnitude and a positive semidefinite Hessian within the class of single-invariant theories.

For the electric branch, \(D_i\neq 0\) and \(H_i=0\). At the level of branchwise stationarity and Hessian positivity, Eq.~(\ref{CR3}) is realized only if
\[
d_0=\frac{2}{\eta}\ln\!\left(-\frac{1}{\lambda}\right), \qquad \lambda<-1,\qquad \eta<0.
\]
This parameter region is incompatible with the boundedness condition in Eq.~(\ref{eq:exp_bounded}), which requires \(\lambda>0\) and \(\eta>0\). Furthermore, the general discussion below shows that these electric branch configurations fail to define local minima of the full effective Hamiltonian. Hence, in the exponential model as well, physically admissible Lorentz-violating vacua are realized exclusively in the magnetic branch.

Table~\ref{tab:models_ssb} collects the main results for the magnetic branch in the three models analyzed in this work. For each case, we display the explicit form of the Pleba\'nski potential, the parameter region where the effective Hamiltonian is bounded from below, the subset where this boundedness coexists with spontaneous Lorentz symmetry breaking, and the corresponding vacuum value $h_0$.

\begin{table*}[h]
\renewcommand{\arraystretch}{1.45}
\begin{tabular}{|c|c|c|c|}
\hline
Model $\hat{V}(P)$ 
& $\mathcal H_{\rm eff}$ bounded below 
& Boundedness $+$ Magnetic SSB 
& $h_0$ \\
\hline

$\displaystyle -P+\frac{\lambda P^{3}}{1+\eta P^{2}}$
&
$\displaystyle \eta>0,\quad \eta\ge \frac{9}{8}\lambda$
&
$\displaystyle \frac{9}{8}\lambda \le \eta <
\left(\frac{25}{16}+\frac{5\sqrt{5}}{16}\right)\lambda$
&
\(h_0\), the positive real solution of Eq.~(\ref{eq:rational_h0_polynomial}) \\
\hline

$\displaystyle -P-\frac{\lambda}{\eta}\ln(1+\eta P)$
&
$\displaystyle \lambda\ge 0,\quad \eta>0$
&
$\displaystyle \lambda>8,\quad \eta>0$
&
$h_0=$$\displaystyle \frac{\lambda-2+\sqrt{\lambda(\lambda-8)}}{\eta}$ \\
\hline

$\displaystyle -P-\frac{\lambda}{\eta}\bigl(1-e^{-\eta P}\bigr)$
&
$\displaystyle \lambda > 0,\quad \eta > 0$
&
$\displaystyle \lambda>\frac{e^{3/2}}{2},\quad \eta>0$
&
$h_0=$$\displaystyle \frac{1-2W_{-1}\!\left(-\frac{e^{1/2}}{2\lambda}\right)}{\eta}$ \\
\hline
\end{tabular}
\caption{Summary of the three explicit $\hat{V}(P)$ models considered in this work.}
\label{tab:models_ssb}
\small
\end{table*}

\begin{figure*}[t]
    \centering
    \includegraphics[width=0.95\textwidth]{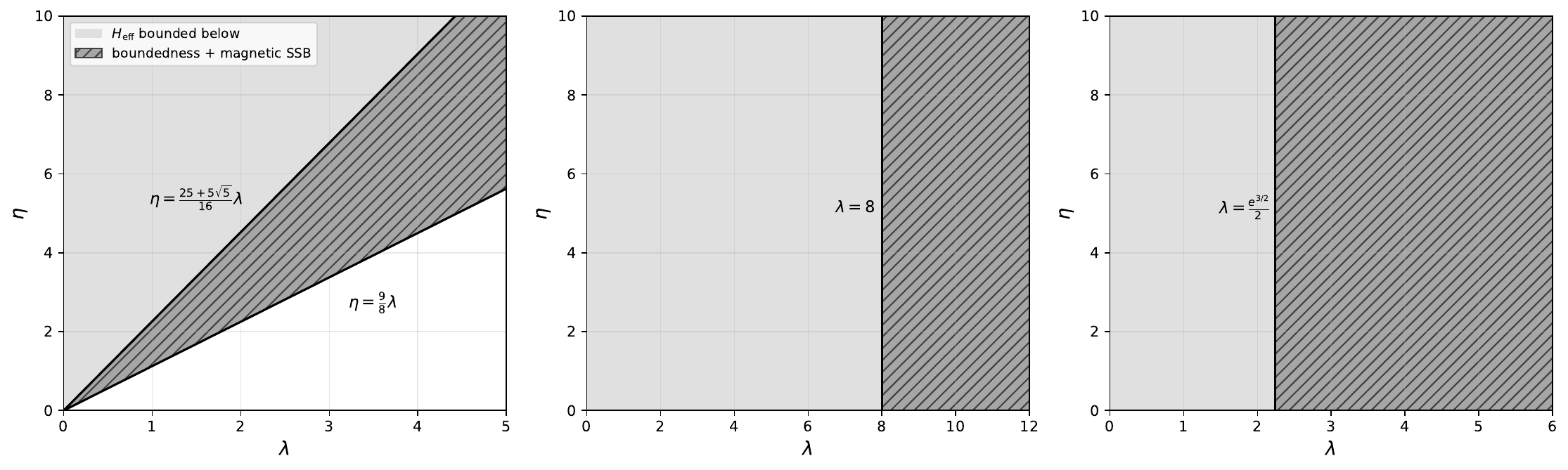}
    \caption{
    Parameter regions for the three \(\hat V(P)\) models. 
    Light-gray regions denote boundedness of the effective Hamiltonian, while hatched regions indicate coexistence of boundedness with magnetic spontaneous Lorentz symmetry breaking.
    }
    \label{fig:parameter_regions}
\end{figure*}

A graphical summary of the admissible parameter regions is shown in Fig.~\ref{fig:parameter_regions}. 
The light-gray regions denote the domains where the effective Hamiltonian is bounded from below, while the hatched regions indicate the corresponding subsets where boundedness coexists with magnetic spontaneous Lorentz symmetry breaking. 
This representation also makes clear that the three parameter planes should not be compared directly: the mass dimensions of \(\lambda\) and \(\eta\) are model dependent. 
Indeed, for the rational asymmetric model one has \([\lambda]=[\eta]=[P]^{-2}\), whereas for the logarithmic and exponential models \([\lambda]=1\) and \([\eta]=[P]^{-1}\).

\subsection{General no-go result for electric vacua}

A common outcome of the three two-parameter models analyzed above is that the electric branch never yields a physically admissible Lorentz-violating vacuum. In the explicit examples considered here, the electric branch is incompatible with the globally bounded region of the effective Hamiltonian. More generally, however, the obstruction does not rely on the specific form of the rational, logarithmic, or exponential nonlinearities, but follows directly from the structure of the effective Hamiltonian in the electric branch.

Consider an arbitrary theory of the form \(\hat{V}(P)\) in the Pleba\'nski first-order formulation, with effective Hamiltonian
\[
\mathcal H_{\rm eff}(h,d)=-h\,\hat V_P(P)+\hat V(P), \qquad P=\frac12(h-d).
\]
Let \((h,d)=(0,d_0)\) be an electric stationary configuration, so that
\[
P_0=-\frac{d_0}{2}<0, \qquad \hat V_P(P_0)=0.
\]
At the level of branchwise stationarity and Hessian positivity, Eq.~(\ref{CR3}) requires that the only nonvanishing Hessian eigenvalue satisfy
\[
\mu_1=d_0\,\hat V_{PP}(P_0)>0,
\]
that is, \(\hat V_{PP}(P_0)>0\).

However, this condition is not sufficient to guarantee a local minimum of the full effective Hamiltonian. Keeping \(d=d_0\) fixed and turning on a small magnetic perturbation \(h>0\), one has
\[
P=P_0+\frac{h}{2},
\]
and therefore
\[
\mathcal H_{\rm eff}(h,d_0)=\mathcal H_{\rm eff}(0,d_0)-\frac{3}{8}\hat V_{PP}(P_0)h^2+O(h^3).
\]
Hence, whenever \(\hat V_{PP}(P_0)>0\), the leading correction is negative, so the energy decreases for sufficiently small \(h>0\). Equivalently,
\[
\left.\frac{\partial \mathcal H_{\rm eff}}{\partial h}\right|_{(0,d_0)}=0,
\qquad
\left.\frac{\partial^2 \mathcal H_{\rm eff}}{\partial h^2}\right|_{(0,d_0)}
=-\frac34 \hat V_{PP}(P_0)<0.
\]
Thus, the electric stationary point is necessarily unstable along the magnetic direction.

Since \(h=\vec H\cdot \vec H\), this instability arises only at quartic order in the magnetic fluctuations, which explains why it is not detected by the quadratic Hessian analysis in the canonical variables. The vanishing magnetic block in the electric Hessian therefore does not correspond to a harmless flat direction, but rather hides a higher-order instability of the full effective Hamiltonian.

This yields a general no-go result for physically admissible electric vacua in single-invariant Pleba\'nski theories $\hat V(P)$. More precisely, an electric stationary configuration satisfying
\[
\hat V_P(P_0)=0, \qquad \hat V_{PP}(P_0)>0, \qquad P_0<0,
\]
cannot define a physically admissible Lorentz-violating vacuum, because it is necessarily destabilized by sufficiently small magnetic perturbations.

\subsection{Additional single-parameter \(\mathcal{L}(\mathcal{F})\) models without spontaneous symmetry breaking}
\label{NOSSB}
 To further illustrate that the coexistence of boundedness, local stability, and spontaneous Lorentz symmetry breaking is not generic, we briefly consider several additional single-parameter \(\mathcal{L}(\mathcal{F})\) models, namely the exponential model introduced by Kruglov \cite{Kruglov}, as well as Born--Infeld-type logarithmic, exponential and square-root models discussed in the literature \cite{NLED}. Although some of them possess Hamiltonians that are bounded from below---either globally or within a restricted physical domain---none of them admit nontrivial vacuum configurations satisfying the symmetry-breaking conditions discussed in Sec.~\ref{Hamiltonian1}. Since the additional examples are usually specified in the standard $\mathcal{L}(\mathcal{F})$ formulation, some care is required in applying the Hamiltonian criteria developed above. The relevant object for the present analysis is the Pleba\'nski potential $\hat V(P)$ associated with each model through the Legendre map. Thus, each entry in Table~\ref{tab:extra_models} is understood together with its corresponding $\hat V(P)$,
obtained parametrically from
\[
P(\mathcal{F})=\mathcal{F} \mathcal{L}_\mathcal{F}^2,
\qquad
\hat V(\mathcal{F})=-\mathcal{L}(\mathcal{F})+2\mathcal{F} \mathcal{L}_\mathcal{F}.
\]

One may then either analyze the effective Hamiltonian directly in terms of $\hat V(P)$, or translate the magnetic vacuum condition to the $\mathcal{L}(\mathcal{F})$ variables. On regular Legendre branches,
\begin{equation}
S_m=\hat V_P+2P\hat V_{PP}
=\frac{1}{\mathcal{L}_\mathcal{F}+2\mathcal{F} \mathcal{L}_{\mathcal{F}\mathcal{F}}} .
\end{equation}

Consequently, a magnetic SSB vacuum would require \(S_m=0\) at some \(P>0\). In the \(\mathcal{L}(\mathcal{F})\) variables this condition must be interpreted through the corresponding regular or limiting behavior of the Legendre map. For the models listed below no such magnetic vacuum is found. The results are summarized in Table \ref{tab:extra_models}.

\begin{table*}[h]
\centering
\renewcommand{\arraystretch}{2.2}
\begin{tabular}{|c|c|c|c|c|}
\hline
Model & \(\mathcal{L}(\mathcal{F})\) & Allowed domain & $\mathcal{H}_{\textrm{eff}}$ bounded below & Magnetic SSB  \\
\hline
(a) &
\(-\mathcal{F}\,e^{-\beta \mathcal{F}}\) &
\(\mathcal{F}\in\mathbb{R}\) &
\(\times\) &
\(\times\)  \\
\hline
(b) &
\(4\beta^{2}\!\left(1-\sqrt{1+\dfrac{\mathcal{F}}{2\beta^{2}}}\right)\) &
\(1+\dfrac{\mathcal{F}}{2\beta^{2}}\ge 0\) &
\(\checkmark\) (restricted domain) &
\(\times\) \\
\hline
(c) &
\(\beta^{2}\!\left(e^{-\mathcal{F}/\beta^{2}}-1\right)\) &
\(\mathcal{F}\in\mathbb{R}\) &
\(\checkmark\) &
\(\times\)   \\
\hline
(d) &
\(-8\beta^{2}\ln\!\left(1+\dfrac{\mathcal{F}}{8\beta^{2}}\right)\) &
\(1+\dfrac{\mathcal{F}}{8\beta^{2}}>0\) &
\(\checkmark\) (restricted domain) &
\(\times\)  \\
\hline
\end{tabular}
\caption{Additional nonlinear \(\mathcal{L}(\mathcal{F})\) models without spontaneous symmetry breaking. The boundedness entries refer to the effective Hamiltonian constructed from the corresponding Pleba\'nski potential.}
\label{tab:extra_models}
\end{table*}

These examples show that boundedness of the effective Hamiltonian, even when present, is not sufficient to guarantee spontaneous Lorentz symmetry breaking. The stationarity and stability conditions impose additional nontrivial restrictions on the nonlinear potential.

A noteworthy difference with respect to the additional models in Table \ref{tab:extra_models} is that the three main examples studied in this work depend on two independent parameters. This allows for a broader adjustment of the theory, since the requirements of boundedness and spontaneous Lorentz symmetry breaking need not be controlled by a single scale. By contrast, in the one-parameter models discussed above, the available freedom is more restricted, which severely limits the possibility of simultaneously accommodating the different physical conditions of interest.

A related conclusion was reached previously for ModMax electrodynamics, originally introduced by Bandos et al.~\cite{Bandos2020} and further developed by Russo and Townsend \cite{T1}. Since ModMax depends on both invariants \(\mathcal{F}\) and \(\mathcal{G}\), it lies outside the class of \(\mathcal{L}(\mathcal{F})\) models considered in Table~\ref{tab:extra_models}. Even so, its Hamiltonian analysis in the Pleba\'nski first-order formulation showed that the theory is bounded from below and yet does not exhibit spontaneous Lorentz symmetry breaking \cite{Mod12}. This provides an additional indication that, in nonlinear electrodynamics with constraints, boundedness of the Hamiltonian by itself does not guarantee the existence of nontrivial Lorentz-violating vacua.

\subsection{Causality and wave propagation}
\label{Causal123}

A further point concerns the relation between the symmetry-breaking vacua identified above and the causality criteria discussed in the literature. For nonlinear electrodynamics of the Pleba\'nski class, Schellstede, Perlick, and L\"ammerzahl showed that, for one-invariant theories of the form \(\mathcal{L}(\mathcal{F})\), causality is governed by the inequalities given in Eqs.~(65) and (66) of Ref.~\cite{Lamm}. In the parity-preserving formulation used by Russo and Townsend, the corresponding strong-field causality condition is written as Eq.~(2.15) in Ref.~\cite{Townsend}. These two conditions are equivalent once one restricts to single-invariant theories and translates between the corresponding sets of variables.

In the Hamiltonian formulation, Ref.~\cite{Lamm} rewrites the strong-field causality condition for \(H(R)\)-theories in Eq.~(80). For the vacuum branches considered here, where the second invariant vanishes, this condition takes a particularly simple form. Identifying the Hamiltonian variable \(R\) with the Pleba\'nski invariant \(P\), one finds that for the magnetic branch (\(P>0\)) the strong-field causality condition reduces to
\begin{equation}
\hat V_{P}+2P\,\hat V_{PP}>0,
\end{equation}
whereas for the electric branch (\(P<0\)) it reduces to
\begin{equation}
\hat V_{P}>0.
\end{equation}
These are precisely the combinations that in our notation define the branch-dependent quantities
\begin{equation}
S_{m}=\hat V_{P}+2P\,\hat V_{PP},
\qquad
S_{e}=\hat V_{P}.
\end{equation}

Therefore, the vacuum conditions for spontaneous Lorentz symmetry breaking,
\[
S_m=0 \quad \text{or} \quad S_e=0,
\]
locate the symmetry-breaking vacua exactly at the boundary where the strong-field causality condition is saturated. In this sense, the appearance of nontrivial Lorentz-violating vacua is directly tied to the edge of the causal domain.

This observation also applies to the additional one-parameter models listed in Table~\ref{tab:extra_models}. Since they belong to the class of single-invariant theories \(\mathcal{L}(\mathcal{F})\), the general analysis of Ref.~\cite{Lamm} implies that any genuinely nonlinear member of this class violates causality for some allowed background fields, Maxwell theory being the only exception. Thus, even when some of these models possess an effective Hamiltonian bounded from below, they still belong to a class that is generically obstructed by causality.


A useful geometric interpretation of this saturation is provided by the effective-metric viewpoint of Novello et al.~\cite{Novello2000}. In nonlinear electrodynamics, the \emph{effective metric} is the background-dependent optical metric that governs the characteristic surfaces of electromagnetic disturbances; equivalently, it defines the null cone along which small field discontinuities propagate, replacing the Minkowski light cone of Maxwell theory. For one-invariant nonlinear electrodynamics, this optical metric takes the form
\begin{equation}
g_{\rm eff}^{\mu\nu}=\mathcal{L}_\mathcal{F}\,\eta^{\mu\nu}-4\mathcal{L}_{\mathcal{F}\mathcal{F}}F^{\mu}{}_{\alpha}F^{\alpha\nu},
\end{equation}
so that, for a purely magnetic background and propagation orthogonal to the background field, the extraordinary mode propagates with
\[
v_\perp^2=\frac{\mathcal{L}_\mathcal{F}+2\mathcal{F} \mathcal{L}_{\mathcal{F}\mathcal{F}}}{\mathcal{L}_\mathcal{F}}.
\]
Passing to the Pleba\'nski formulation, this expression can be rewritten in terms of the Hamiltonian potential as

\[
v_\perp^2=\frac{\hat V_P}{\hat V_P+2P\,\hat V_{PP}}.
\]
In the magnetic branch considered here, where \(2P=h\), this becomes
\[
v_\perp^2=\frac{\hat V_P}{\hat V_P+h\hat V_{PP}}=\frac{\hat V_P}{S}.
\]

Therefore, as $S\to 0$ with $\hat{V}_P\neq 0$, the propagation speed becomes unbounded, showing that the symmetry-breaking condition does not merely saturate the strong-field causality inequality, but corresponds to a degeneration of the effective causal cone itself.

This result also suggests a close connection with the degenerate behavior discussed in Ref.~\cite{Degbeha}. There it was shown, within the Pleba\'nski first-order formulation, that hypersurfaces on which quantities such as \(\hat V_{P}\) or \(\hat V_{P}+2P\hat V_{PP}\) vanish may lead to singular behavior in the equations of motion, including the loss of local degrees of freedom and shock-wave-like evolution. In the explicit example analyzed in Ref.~\cite{Degbeha}, one of the propagation modes becomes superluminal as the background field approaches the degenerate hypersurface. From this perspective, the present symmetry-breaking vacua, characterized by \(S_{m}=0\) or \(S_{e}=0\), should be viewed as lying precisely on the edge of a region where causal propagation may fail, consistently with the strong-field causality criteria of Refs.~\cite{Lamm,Townsend}.

\section{Conclusions}
\label{Conclu}

In this work we have carried out a comparative Hamiltonian analysis of several gauge-invariant nonlinear electrodynamics models within the Pleba\'nski first-order formulation. The main outcome is that the simultaneous realization of global boundedness of the effective Hamiltonian, local stability of the vacuum, and spontaneous Lorentz symmetry breaking is highly nontrivial within the class of single-invariant theories.

Our results show that the three two-parameter models studied in detail---namely the rational asymmetric, logarithmic, and exponential models---admit physically admissible Lorentz-violating vacua only in the magnetic branch. By contrast, the electric branch does not define physically admissible vacua of the full effective Hamiltonian. In the explicit models considered here, the electric branch is incompatible with global energetic admissibility, and more generally it is destabilized by arbitrarily small magnetic perturbations. This asymmetry between magnetic and electric vacua is one of the most robust features that emerges from the present analysis, and suggests that, within the corresponding single-invariant class, the Hamiltonian constraints strongly favor magnetic symmetry-breaking backgrounds.

A second conclusion is that boundedness of the effective Hamiltonian, even when it holds globally or within a restricted physical domain, is not sufficient to ensure spontaneous Lorentz symmetry breaking. This is made explicit by the additional one-parameter models considered in Sec.~\ref{NOSSB}, none of which exhibits nontrivial symmetry-breaking vacua. In this sense, the stationarity and Hessian conditions play a genuinely selective role, and the successful realization of spontaneous Lorentz violation requires a rather specific structure of the nonlinear potential. The comparison with the one-parameter examples also indicates that the additional freedom provided by two independent parameters is useful in reconciling the different requirements imposed by boundedness and local stability.

The analysis also points to a close connection between spontaneous symmetry breaking and causal propagation in nonlinear electrodynamics. As discussed in Sec.~\ref{Causal123}, the branch-dependent quantities \(S_m\) and \(S_e\) that characterize the Lorentz-violating vacua are precisely the combinations that appear in the strong-field causality criteria for single-invariant theories \cite{Lamm,Townsend}. Therefore, the conditions \(S_m=0\) and \(S_e=0\) place the symmetry-breaking vacua at the boundary where the causal inequalities are saturated. From this viewpoint, the vacua found here should be regarded not as generic interior points of a causal region, but as distinguished limiting configurations of the theory.

This interpretation is reinforced by the degenerate behavior previously identified in the Pleba\'nski framework \cite{Degbeha}. There it was shown that hypersurfaces defined by the vanishing of combinations such as \(\hat V_P\) or \(\hat V_P+2P\hat V_{PP}\) may lead to singular behavior in the equations of motion, including a reduction in the number of local degrees of freedom and shock-wave-like evolution. Moreover, explicit examples show that superluminal propagation may arise as such a hypersurface is approached. In this sense, the present Lorentz-violating vacua lie at the interface between vacuum structure and propagation pathologies, which makes their interpretation more subtle and, at the same time, more interesting.

These results naturally suggest several extensions. A first direction is to go beyond one-invariant models and consider theories of the form \(\mathcal{L}(\mathcal{F},\mathcal{G})\), or parity-preserving subclasses such as \(\mathcal{L}(\mathcal{F},\mathcal{G}^2)\), where the additional invariant may relax the tension between boundedness, symmetry breaking, and causality. A second direction is to analyze the optical metrics and Fresnel structure associated with the vacua found here, in order to determine whether birefringence, cone deformation, or loss of hyperbolicity occur in their vicinity. A third direction is to investigate whether similar symmetry-breaking branches can exist in special families such as self-dual or no-birefringence nonlinear electrodynamics. Finally, it would be worthwhile to couple these models to gravity and examine whether the Lorentz-violating vacua remain meaningful in self-gravitating configurations.

Overall, the present work shows that the Pleba\'nski Hamiltonian formulation provides a useful framework not only for constructing explicit nonlinear electrodynamics models with spontaneous Lorentz symmetry breaking, but also for identifying the restrictions imposed by stability and causality. From this perspective, the models analyzed here should be viewed as part of a broader program aimed at classifying those gauge-invariant nonlinear electrodynamics theories for which nontrivial vacua can exist without losing control of the Hamiltonian and causal structure.

\begin{acknowledgments}
R. L. acknowledges partial support from CONAHCyT-M\'exico under Grant No. CBF-2023-2024-1937.
EPF would like to thank Secretar\'ia de Ciencias, Humanidades, Tecnolog\'ia e Innovaci\'on, SECIHTI (M\'exico) and Universidad Aut\'onoma Metropolitana-Iztapalapa (UAM-I) for their continued support.
\end{acknowledgments}


\begin{thebibliography}{99}


\bibitem{KosteleckySamuel1989} V.A. Kosteleck\'y and S. Samuel, Phys. Rev. D \textbf{40}, 1886 (1989). 

\bibitem{Amelino} G. Amelino-Camelia, J.R. Ellis, N.E. Mavromatos, D.V. Nanopoulos, and S. Sarkar, Nature \textbf{393}, 763 (1998).

\bibitem{Kostelecky} D. Colladay and V. A. Kosteleck\'y, Phys. Rev. D \textbf{55}, 6760 (1997).

\bibitem{Kostelecky2} D. Colladay and V. A. Kosteleck\'y, Phys. Rev. D \textbf{58}, 116002 (1998).

\bibitem{Hernaski1} C. A. Hernaski, Phys. Rev. D \textbf{90}, 124036 (2014).

\bibitem{Escobar1} C. A. Escobar and A. Mart\'in-Ruiz, Phys. Rev. D \textbf{95}, 095006 (2017).

\bibitem{Potting1} R. Bluhm, N. Gagne, R. Potting and A. Vrublevskis, Phys. Rev. D \textbf{77}, 125007 (2008).


\bibitem{Kostelecky2004} V. A. Kostelecký, Phys. Rev. D \textbf{69}, 105009 (2004).


\bibitem{Altschul2010} B. Altschul, Q. G. Bailey, and V. A. Kostelecký, Phys. Rev. D \textbf{81}, 065028 (2010).

\bibitem{Gomes2008} M. Gomes, T. Mariz, J. R. Nascimento, and A. J. da Silva, Phys. Rev. D \textbf{77}, 105002 (2008).

\bibitem{Assuncao2017} J. F. Assunção, T. Mariz, J. R. Nascimento, and A. Yu. Petrov, Phys. Rev. D \textbf{96}, 065021 (2017).

\bibitem{Potting2024} R. Potting, Phys. Rev. D \textbf{110}, 045014 (2024).

\bibitem{Assuncao2019}
J. F. Assun\c{c}\~ao, T. Mariz, J. R. Nascimento and A. Yu. Petrov,
Phys. Rev. D \textbf{100}, 085009 (2019).

\bibitem{Ferrari2007}
A. F. Ferrari, M. Gomes, J. R. Nascimento, E. Passos, A. Yu. Petrov and A. J. da Silva,
Phys. Lett. B 652, 174 (2007).

\bibitem{Gomes2010}
M. Gomes, J. R. Nascimento, A. Yu. Petrov and A. J. da Silva,
Phys. Rev. D 81, 045018 (2010).

\bibitem{Mariz2018}
T. Mariz, R. V. Maluf, J. R. Nascimento and A. Yu. Petrov,
Int. J. Mod. Phys. A \textbf{33}, 1850018 (2018).

\bibitem{Scarpelli2013}
A. P. Baeta Scarpelli, T. Mariz, J. R. Nascimento and A. Yu. Petrov, Eur. Phys. J. C \textbf{73}, 2526 (2013).


\bibitem{Boillat1970} G. Boillat, J. Math. Phys. \textbf{11}, 941 (1970).

\bibitem{Novello2000} M. Novello, V. A. De Lorenci, J. M. Salim, and R. Klippert, Phys. Rev. D \textbf{61}, 045001 (2000).

\bibitem{ObukhovRubilar2002} Y. N. Obukhov and G. F. Rubilar, Phys. Rev. D \textbf{66}, 024042 (2002).

\bibitem{GibbonsRasheed1995} G. W. Gibbons and D. A. Rasheed, Nucl. Phys. B \textbf{454}, 185 (1995).

\bibitem{Pleb} J. Pleba\'nski, \textit{Lectures on non-linear electrodynamics}, (Copenhagen: Nordita) 1970.

\bibitem{EscobarPotting2020}C. A. Escobar and R. Potting, Int. J. Mod. Phys. A \textbf{35}, 2050174 (2020).

\bibitem{Dirac1} P.A.M. Dirac, \textit{Lectures on Quantum Mechanics}, (Dover Publications, New York, 2001).


\bibitem{Kruglov} S. I. Kruglov, EPL \textbf{115}, 60006 (2016).

\bibitem{NLED} S. H. Hendi, J. High Energ. Phys. \textbf{2012},65 (2012). 



\bibitem{Bandos2020} I. Bandos, K. Lechner, D. Sorokin, and P. K. Townsend,
Phys. Rev. D \textbf{102}, 121703 (2020). 

\bibitem{T1} J. G. Russo and P. K. Townsend,
J. High Energy Phys. \textbf{01}, 039 (2023).


\bibitem{Mod12} C. A. Escobar, R. Linares, and B. Tlatelpa-Mascote,
Int. J. Mod. Phys. A \textbf{37}, 2250011 (2022).


\bibitem{Lamm} G. O. Schellstede, V. Perlick and C. Lammerzahl, Annalen Phys. \textbf{528}, 738-749 (2016).

\bibitem{Townsend} J.G. Russo and P. K. Townsend, J. High Energ. Phys. \textbf{06}, 191 (2024). 

\bibitem{Degbeha} C. A. Escobar and R. Potting, Phys. Scr. \textbf{95}, 065218 (2020).

\end{thebibliography}
\end{document}